\documentstyle[12pt]{article}
\begin{document}

\hfill {WM-94-101}

\hfill {January 1994} \vskip 1in  \baselineskip 24pt

\begin{centering}
           {\large Quark cluster signatures
              in deuteron electromagnetic interactions}
  \vskip .5in

C.E. Carlson \bigskip

{\it Physics Department, College of William and Mary

Williamsburg, VA 23187, USA}  \bigskip

K.E. Lassila \bigskip

{\it Department of Physics and Astronomy

Iowa State University,  Ames,  IA 50011,  USA}

\end{centering}  \vskip 1in

{\narrower \narrower Abstract. A suggestion is made for
distinguishing 2N and 6q short range correlations within the
deuteron.  The suggestion depends upon observing high momentum
backward nucleons emerging from inelastic electromagnetic scattering
from a deuteron target.  A simple model is worked out to see the
size of effects that may be  expected.

}
\newpage

\section{Introduction}

	In deep inelastic scattering upon deuterons or heavier nuclei,
nucleons or other hadrons  can emerge backward to the direction
defined by the incoming photon in the target rest frame.   The
backward nucleons do not,  we believe,  come from the nucleon or the
quark that was  struck \cite{fs79}.  Rather,  the backward nucleons
come from the debris that remains after the item that  was struck is
driven strongly forward.  We have commented \cite{cls91,cl93} upon
using neutrino  production of backward protons
\cite{mats89,ammo86,efre80,berg78} to explore short distance quark
configurations and here wish to use deep inelastic
electromagnetic interactions to similar ends.

	We distinguish two extreme cases. In a 2-nucleon or 2N correlation
the nucleons maintain their characters as neutrons and protons no
matter what their separation is.  To get a backward proton from
the deuteron,  one must strike the neutron.  This will break up  the
bound state,  and the proton will emerge with the Fermi momentum it
has at the moment of  breakup,  and the Fermi momentum will be
backwards in the case of interest.

	The other extreme case is the 6-quark or 6q cluster.  Here we mean a
``kneaded'' 6q,   $uuuddd$,  object with all the quarks in relative
S-states.  The flavor-spin-color wave function  is unique (for
overall deuteron quantum numbers) and is not equivalent to two
nucleons lying  at the same spatial point.  Emitting a backward
proton begins with one quark being struck and  driven forward.  The
proton must be formed out of the remaining 5 quarks,  plus possible
higher Fock components,  and the process of forming hadrons we refer
to as the  ``fragmentation'' of the 5-quark residuum.  The
term fragmentation follows common usage for the  production of
hadrons from any color non-singlet QCD object, quark and gluon
jets being the most familiar. In the breakup of this 5-quark
residuum,  we persist with the nomenclature ``fragmentation''
although recombination may be the process chiefly at work.  In any
case,  the 6q model can produce a backward proton spectrum which
agrees with data
from neutrino reactions,  for backward hemisphere proton momentum
above about 300 MeV
\cite{cls91}.  However,  so can the 2N model with enhanced high
momentum components \cite{atti91}.  We need a more detailed
indicator to test the two models.  For weak interactions,  ratios
of neutrino and antineutrino induced backward proton production
cross sections cancel much of what is unknown in either the 2N or
6q model and yet give predicted results which are not the same for
the two models \cite{cl93}.  Similar opportunities should also
exist for electromagnetic interactions.

	Here we suggest a characteristic in the spectrum of backward
nucleons which is reliably different for the two scenarios of the
short range configuration,  and which allows electromagnetic
experiments to adjudicate between them.  A suggestion for a ratio to
examine at fixed backward proton momentum and varying Bj\"ork\'en $x$
is given in section 2,  along with some numerical estimates of the
differences between the two models.  Further comments on
possibilities at CEBAF,  Fermilab,  or elsewhere are made in section
3.  Incidentally,  an observation regarding changes in average
Bj\"ork\'en $x$ with varying backward nucleon momentum,  which was
originally made in the context of the 2N model,  should be
reasonably true for most any model and we also comment on this in
section 3.  We conclude in section 4.

\section{ Signatures of short distance correlations}

\subsection{\it General}

	The cross section for inelastic electromagnetic scattering of a
lepton from a stationary  target is (in the scaling region and
neglecting $\sigma_L/\sigma_T$)
\begin{equation}{{d\sigma } \over {dx\,dy}}\;=\;{{4\pi \,\alpha_{em}
^2\,m_N\,E} \over {Q^4}}\;\left( 1+\left(  {1-y} \right)^2
\right)\;F_2(x,Q^2)\end{equation}
	where $E$ and $E^\prime$ are the incoming and outgoing lepton
energies,
$\nu$ is the difference  between them,  $y = \nu/E$,  $x =
Q^2/2m_N\nu$,  and $F_2$ (whose $Q^2$ dependence will generally be
tacit) is
 \begin{equation}F_2(x)\;=\;\sum\limits_{}
{}e_i^2\,x\,q_i(x)\end{equation}  where $e_i$ is the quark charge
in units of proton charge and $q_i$ is the distribution function for
a  quark of flavor $i$.

\subsection{\it Backward nucleons from 2N correlations}

	We will speak of observing a backwards proton for the sake of
definiteness;  observing a backwards neutron is essentially similar.

	Some things change in the above formula when a backwards proton is
observed.  The neutron,  which is the struck  particle,  is not at
rest in the lab frame.  Then the momentum fraction of the struck
quark relative to the neutron is not $x$ but rather $\xi$,
 \begin{equation}\xi \;=\;{x \over {2-\alpha }},\end{equation}
where $\alpha = (E_p + p^z) /m_N$ is the light cone momentum
fraction of the backward proton with $p^z$ positive for
a backward proton.  One should also replace $m_N E$ by
its corresponding Lorentz invariant,
\begin{equation}m_N E\;\to \;p_n\cdot k\;=\;m_N E (2-\alpha).
\end{equation}

 Also,  if we want a backward proton,  there is a further factor of
the probability density  for finding the proton with its observed
momentum at the moment the neutron was struck.   Thus
\begin{equation}\sigma _{2N}\;\equiv \;{{d\sigma _{2N}} \over
{dx\,dy\,d\alpha
\,d^2p_T}}\;=\;{{4\pi
\,\alpha_{em}^2\,m_N\,E} \over {Q^4}}\;\left( {1+\left( {1-y}
\right)^2}
\right)\;(2-\alpha  )\,F_{2n}(\xi )\cdot \left| {\psi (\alpha ,p_T)}
\right|^2,\end{equation}
 where $p_T$ is the transverse momentum of the backward proton and
we used the light cone wave  function normalized by
\begin{equation}\int {}d\alpha \,d^2p_T\;\left| {\psi (\alpha ,p_T)}
\right|^2\;=\;1.\end{equation}

	The neutron structure function $F_{2n}$ is (we shall suppose)
known.  The test we propose is to measure the cross section for
backward nucleon production at a variety of $x$ and $\alpha$ and
examine the ratio
\begin{equation}R_1\;=\;{{{{\sigma _{meas}} \mathord{\left/
{\vphantom {{\sigma _{meas}} K}} \right.
\kern-\nulldelimiterspace} K}} \over {F_{2n}(\xi )}}.\end{equation}
Here, $\sigma_{meas}$ is the measured differential cross  section
and $K$ is the factor

\begin{equation}K\;=\;{{4\pi \,\alpha_{em}^2\,m_N\,E} \over
{Q^4}}\;\left( {1+\left( {1-y}
\right)^2} \right) .\end{equation}

The signature of a two nucleon correlation model is that this ratio
is independent of $x$ for any fixed $\alpha$ and $p_T$.  Of course,
how useful this signature is depends upon how different we may
expect the result to be for a 6q cluster.  This we shall see in the
next section.

\subsection{\it Backward nucleons from 6q clusters}

	For the case of electromagnetic scattering from the 6q state,  we
have basically the  convolution of $F_2^{(6)}$ with the
fragmentation functions of the five (or more,  in general)  quark
residuum.  Since the quarks in the residuum depend on which flavor
quark was struck,   we must write
\begin{equation}\sigma _{6q}\;\equiv \;{{d\sigma _{6q}} \over
{dx\,dy\,d\alpha
\,d^2p_T}}\;=\;K\;\sum\limits_i {}x\,e_i^{\,2}V_i^{(6)}(x)\cdot {1
\over {2-x}}\;D_{p/  5q}(z,p_T).\end{equation}
Here $V_i^{(6)}$ is the distribution  function
for a valence quark in a six quark cluster,  the sum is over quark
flavors,  and $D_{p/5q}$  is the fragmentation function for the 5q
residuum,  i.e.,  the probability density per unit $z$ and $p_T$  for
finding a proton coming from the 5q cluster.  It is tacit that the
correct 5q cluster,  either  $u^2d^3$ or $u^3d^2$,  is chosen.
Argument $z$ is the fraction of the residuum's light-cone
longitudinal momentum that goes into the proton,
\begin{equation}z\;=\;{\alpha  \over {2-x}};\end{equation}
 the factor $(2-x)^{-1}$ comes because $D_{p/5q}$ is probability per
unit $z$ in its definition and we  quote the differential cross
section per unit $\alpha$.  The formula is written for high
backward proton momentum,  where we can expect the  5q residuum
and hence the actual 6q initial state to dominate.

Neither
$F_2^{(6)}$ nor $D_{p/5q}$ can be said to be  known.  However,
since a large fraction of the short range part of the baryon number
two state may be in a 6-quark cluster,  we should make the best
guess as to what  these functions might be and see how large a
difference it could make experimentally to have significant 6q
cluster contributions,  at least in given regions of phase space.

	Estimates of $F_2^{(6)}$  in a model where the nuclei are treated
as containing some fraction 6q clusters have been given
by Lassila and Sukhatme \cite{ls88}.  They chose their quark
distributions beginning with quark counting rules and then fine
tuned with physical logic to describe the EMC data.  For
completeness,  the three parameterizations they present are recorded
in the  Appendix.

	The fragmentation function is even less well known since there is no
complete body of  data to check it against.  The counting rules
suggest a cubic dependence,  as (unnormalized)
      \begin{equation}D_{p\// q^5}(z)\;\propto
\;(1-z)^3\end{equation}      for $z \rightarrow  1$ and for zero
$p_T$ .  We shall use this form,  although we keep in mind the
possibility  that higher order contributions or renormalization
group considerations could somewhat alter  the power,  as they do in
many parameterizations of the quark distributions in nucleons.

	If the high momentum backward protons come from a 6-quark cluster,
then
$\sigma_{meas} =  \sigma_{6q}$ and the experimental ratio $R_1$
should be given by
\begin{equation}R_1\;=\;R_1^{(6)}\;=\;{{F_2^{(6)}(x)\;D_{p\//
5q}(z,p_T)} \over {(2-x)\,(2-\alpha  )\;F_{2n}(\xi
)}}.\end{equation}
 There is no reason for $R_1^{(6)}$ to be independent of $x$ for
fixed
$\alpha$ and
$p_T$ .  We plot this $R_1$ in  Figs. \ref{fig1} and \ref{fig2} for
$p_T$ = 0 and specified
$\alpha$.  Some old and simple quark distributions of Carlson  and
Havens \cite{ch83} were used to get $F_{2n}$ in Fig. \ref{fig1} and
the CTEQ1L \cite{cteq93} distributions were used  in Fig.
\ref{fig2}.  The difference between what is seen and the horizontal
line expected from a pure 2N  correlation model is not negligible.
That the ratio goes to a finite value as we reach the  kinematic
limit $x = 2 - \alpha$ in Fig. \ref{fig1} has to do with the fact
that both the 5q fragmentation  function and the dominant quarks in
the nucleon approach their end points cubically in our  models.  A
dive to zero or a flight to infinity is not precluded in real life,
and the latter is seen  in Fig.
\ref{fig2}.

\section{ Commentary}

\subsection{\it Potential dominance of 6q cluster}

We expect that a correct description of the deuteron would have
a neutron and a proton treated as in the 2N model when they are far
apart.  As they get closer,  QCD processes such as gluon
recombination \cite{mq86} will surely occur and affect first the
ocean parton distributions.  It is something of a simplification to
think of a deuteron as just a combination with a large fraction pure
2N state with a small fraction (perhaps 5\%,  from wave function
overlap estimates \cite{scpv}) 6q state added in.

However, we emphasize that while a 6q cluster may be a small part
of the deuteron overall,  it could be a large fraction of the short
range part of the  deuteron wave function.  The deviation from what
is expected in a pure 2N correlation could be as large as is shown in
our Figures.  The situation is not like the EMC effect where the
effects  of the 6q cluster are diluted by the mostly
ordinary collection of nucleons in the target.  Here we can select
events in a phase space region to enhance 6q cluster effects.  The
backward  proton is a tag that emphasizes the 6q cluster and---if it
is there at all---it will  dominate the cross section for large
enough backward proton momentum.

	It is  necessary to have data at fairly high backward momenta.  No
one doubts that at low relative  momentum or long distances the
deuteron is a neutron plus a proton.  How high is needed?   We
suggest 300 MeV/c is a good starting point,  based on existing
backward proton data from  deep inelastic neutrino scattering and
studies of the backward proton spectrum in that process using 6q
cluster ideas.

\subsection{\it Falling $\langle x\rangle$ with increasing $\alpha$}

Let us point out a piece of kinematics.  From momentum conservation
one has $0 \le x \le (2-\alpha)$.  Hence,  unless the
$x$ distribution has a bizarre shape,  one expects that $\langle x
\rangle_\alpha$ ---the average value of $x$ at fixed
$\alpha$---decreases as
$\alpha$ increases.  This was pointed out in Ref.\ \cite{fs79} in the
context of the 2N model,  and was initially suggested as a test of
that model.  However,  the result should be produced by any model,
so finding the trend in the data is not startling.

The two-nucleon correlation model does give a specific result that
$\langle x \rangle_\alpha$ falls to zero linearly as $2-\alpha$.  In
contrast,  the six-quark cluster model may or my not fall quite
linearly.   It depends on the specific implementation of the model.

The rest of this subsection attempts to show why the two-nucleon
result for $\langle x \rangle_\alpha$ is independent of the internal
details of that model,  but similar manipulations for other models
do not lead to such definite results.  It has to do with the way the
cross section factors.

For the two-nucleon model, the structure of the formula for the cross
section differential in $x$ and $\alpha$ is,
\begin{equation}P(x,\alpha) = {dN\over dx\,d\alpha}=
f(\xi)\cdot{|\psi(\alpha)|^2
\over 2-\alpha},\end{equation}   where $\xi$ is defined earlier.
An elementary calculation gives
\begin{equation}\langle x \rangle_\alpha =
\,{\int_0^{2-\alpha}dx\,x\,P(x,\alpha)
\over \int_0^{2-\alpha}dx\,P(x,\alpha)}=(2-\alpha)\langle \xi
\rangle\,,\end{equation}  where $\langle \xi \rangle$ is
independent of $\alpha$. We can easily turn this into
\begin{equation} {\langle x \rangle_\alpha
\over \langle x \rangle_{\alpha=1}} = (2-\alpha).
\end{equation}

The sort of result that can be derived in the corresponding way for
the six-quark cluster appears less useful.  We envision one quark
being struck and driven forward,  and the residuum that remains
``fragments'' (or recombines) into a nucleon, that often goes
backward,  plus other stuff.  The differential cross section is
structurally
\begin{equation}P(x,\alpha) = {dN\over dx\,d\alpha}=
{g(x)\over2-x}\cdot D(z={\alpha\over2-x})\,.\end{equation} The
factor $g(x)$ is for the quark knockout and $D(z)$ is the
fragmentation function.  The first is a function of $x$ since it
involves distribution functions of quarks in the six-quark cluster,
and the six-quark cluster is standing still in the lab.  Argument
$z$ is defined earlier.  The neatly derivable result is for average
$\alpha$ at fixed $x$:
\begin{equation}\langle \alpha \rangle_x =
\,{\int_0^{2-x}d\alpha\,\alpha\,P(x,\alpha)
\over \int_0^{2-x}d\alpha\,P(x,\alpha)}=(2-x)\langle z
\rangle\,,\end{equation}     or,
\begin{equation}{\langle \alpha \rangle_x
\over \langle \alpha \rangle_{x=1}} = (2-x).
\end{equation}

This result for the six-quark cluster contribution requires
averaging over all $\alpha$, and while the optimist may expect the
six-quark contributions to dominate at high $\alpha$,  no-one
expects them to do so at moderate $\alpha$.  So, this result
seems untestable.

For $\langle x \rangle_\alpha$ details would have to be worked out
for each special case.  However,  remembering the general result that
$\langle x \rangle_\alpha$ decreases with increasing $\alpha$ and is
zero kinematically when $\alpha =2$ makes it likely that one will
get something like the 2N result.

\subsection{\it Possibilities at CEBAF}

	Can CEBAF with a 4 or 6 GeV beam be useful?  We believe yes for the
6 GeV beam.   The question centers around how much backward momentum
is possible with an incoming  electron of this energy,  and how much
data we can get in the scaling region.

\subsubsection{\it Limits on backward proton momentum}

	The maximum directly backward proton momentum when an electron
scatters from a  deuteron is $(3/4) m_N$,  or 704 MeV,  but this is
for the case of infinite incoming energy.  If the  energy is
finite,  the magnitude of the maximum backward momentum is reduced.
For  example,  with $E = 4$ GeV and $Q^2 = 1$ GeV,  the maximum
directly backward proton momentum   is
		\begin{equation}p^{\,z}\;\le
\;.600\,m_N\;=\;564\;MeV,\end{equation}
 which is still a decent backward momentum.  (For a 6 GeV electron
beam and the same $Q^2$,   the maximum $p^{\,z}$ is 609 MeV.)

	These results were obtained with the help of
\begin{equation}\alpha \,m_N\;\le \;{1 \over
2}\;\left( {m_d+\nu -\sqrt {\nu ^2+Q^2}}
\right)\;\left\{ {1+\sqrt {1-{{4m_N^{\,2}} \over {2m_d\nu
-Q^2+m_d^{\,2}}}}}
\right\},\end{equation}
 and also,  for no transverse momentum,
      \begin{equation}p^{\,z}\;=\;m_N\,{{\alpha ^2-1} \over
{2\alpha }}.\end{equation}

 For $\nu \rightarrow \infty$,  we recover the limits cited,  for
example,  in \cite{cls91}.  Now at finite energy and for a  given
$Q^2$ we maximize the limit on $p^{\,z}$ by maximizing $\nu$.  We
have
\begin{equation}\nu \;=\;E-{{Q^2} \over {4E\sin ^2(\theta /2)}} \;\le
\;E-{{Q^2}
\over {4E}}.\end{equation}
 For $Q^2 = 1$ GeV$^2$ and $E = 4$ GeV,  we get $\alpha \le 1.77$
and $p^{\,z}$ as quoted above.

	Much of the limitation actually comes because fixing $Q^2$ for a
given incoming energy  puts a lower limit on Bjorken $x$.  Since
this is also the momentum fraction of the struck quark  in the lab
frame,  it means that the struck quark is not moving forward as fast
as possible,  and  the residuum is then not moving backwards as fast
as possible.  For the 4 GeV beam and the  present $Q^2$ limit,
$x_{min} = 0.14$.  If we went to infinite energy,  but maintained
this value of $x$,   $\alpha$ would still be limited by
$\alpha \le 2 - x_{min} = 1.86$.

\subsubsection{\it The scaling window}

	Using the formulas with the scaling result for $F_{2n}$ requires
that we be in the scaling  region.  This will squeeze the allowed
range of $x$ to be narrower than the kinematic limits.

	Scaling requires,  at a minimum,  that $Q^2$ be above 1 GeV$^2$ and
that
$W$ (the photon plus  single target nucleon c.m. energy) be above 2
GeV,  out of the resonance region.

	The lower limit on $Q^2$ sets a lower limit to $x$,  since for a
given energy there is a  maximum energy transfer $\nu$ possible.
The maximum $\nu$ comes for backward electron  scattering and leads
to $\nu \le E - Q^2/4E$ or
\begin{equation}x_{min}\;=\;{{4\,E\,Q_{min}^{\,2}} \over
{2m_N\,\left( {4E^2-Q_{min}^{\,2}}
\right)}.}\end{equation}

	The lower limit on W sets an upper limit on $x$.  Applying the
limit to the final state that  comes from striking the neutron in
the 2N correlation model gives
\begin{equation}W^2\;=\;(p_n+q)^2\;=\;m_N^{\,2}+2\,m_N\,\nu
\,(\,2-\alpha
\,)-Q^2\;\ge
\;W_{min}^{\,2},\end{equation}
for a given $\alpha$ of the backward proton and letting
$p_n^2=m_N^2$.  This leads to a limit,  also reached  for 180$^\circ$
scattering, that
\begin{equation}x_{max}\;=\;{{2E\;\left( {2m_NE(2-\alpha
)-(W_{min}^{\,2}-m_N^{\,2})}
\right)}
\over {m_N\,\,\left( {4E^2+(W_{min}^{\,2}-m_N^{\,2})}
\right)}}.\end{equation}

	The curves giving $x_{max}$ and $x_{min}$ are shown in Fig.
\ref{fig3} for
$\alpha = 1.4$.  The scaling region is the region between the two
curves.

	For $\alpha$ = 1.4,  the scaling region includes only a short span
of $x$ for electron energy E =  4 GeV,  but the span increases
greatly for E = 6 or 8 GeV.  The span for 8 GeV is over half of  the
maximum possible at any energy,  and not too much less than could be
got at 15 GeV.  For  increasing
$\alpha$,  which corresponds to increasing velocity of the struck
neutron away from the  photon,  larger energy in the laboratory is
needed to reach the scaling region.

\subsubsection{\it Outside the scaling region}

	Do we need to be in the scaling region?  Our formulas for the 2N
correlation are  simplest to evaluate there,  and we don't know how
to guess at forms for production off the 6q  component,  so our
comparison case is gone.  But, from the 2N
viewpoint,  the  purpose of scattering off the neutron is to free
the proton---nothing more.  If all we want is a  yes/no on the 2N
correlation model,  we could take a measured cross section for
producing backward protons and divide by a cross section
for scattering off a neutron at the  correct values of the incoming
variables,  and see if the result depends only upon $\alpha$ and
$p_T$.

	The disadvantage of doing this may be more practical.  Driving the
struck particles  forward forcefully separates them greatly in
momentum space from the backward proton and reduces  the final state
interactions,  which we have neglected in our discussion.  As we
leave the scaling  region,  it can mean that the energy transferred
to the forward moving particles is low enough  that we need to
rethink our attitude towards final state interactions.

\subsection{\it Possibilities with muons at higher energy}

	At Fermilab,  for example,  experiment E665 scattered 490 GeV muons
from deuterium and xenon targets in a streamer chamber so as to be
able to observe backward charges.  A later version of experiment
E665 has capability to observe backward neutrons,  and uses
carbon,  calcium,  and lead as its heavier targets.  The
$x$  dependence of the ratio $R_1$ remains a good
observable.  There is a clear  advantage in having higher energy as
one does not have to think about the ``scaling window.''   It is
virtually the full possible kinematic range.

	Also there is some rate advantage in measuring backward neutrons
instead of protons.   For the 6q model,  one can work out that if
the 6q valence configuration dominates,  as it  should at high
backward nucleon momenta,  then the ratio of neutron rate to proton
rate should  be 3/2 and be independent of $x$.  (One uses a
combinatoric argument to get ratios of proton and  neutron
production from $u^3d^2$ and $u^2d^3$ residua,  and some discussion
of this appears in \cite{cl93}.)   The result for the 2N model is
different,  since one has
 \begin{equation}{{\sigma _{2N}(\mu \,d\,\to \,\mu \,n\,X)} \over
{\sigma _{2N}(\mu \,d\,\to \,\mu  \,p\,X)}}\;=\;{{F_{2p}(\xi )} \over
{F_{2n}(\xi)}},
\end{equation}
 so that the ratio should vary from about 1 at low $\xi$ to about 4
at high $\xi$.  Thus backward  neutrons are always produced more
copiously,  and this is an interesting additional  observable  if
one has the capability of detecting both flavors of nucleons.

	We might also note that study of backward nucleons from a deuteron
target is a study of  the ``target fragmentation region'' and is
best and most easily carried out in the rest frame of the  target,
i.e.,  the lab.  Data presented in the photon-target c.m. is not
equally useful.

\section{ Conclusion}

	Study of the deuteron should be pursued since
deuterons are our chief source of  information about neutrons and we
should understand this source.  Further,  the behavior of the
deuteron state at short range gives information about the short range
dynamics of strong  interaction QCD.

We have suggested a measurement to learn what the short range wave
function of the deuteron is.  Namely,  examine the shape of the
measured differential cross  section for electroproduction of
backward protons or neutrons from a deuteron target,  and take  its
ratio to what would be expected for deep inelastic scattering from a
free neutron or proton,   respectively.  A high momentum backward
nucleon acts as a tag isolating events where the  initial material
in the deuteron was tightly bunched.  The $x$ dependence or lack of
$x$  dependence of the ratio is a signal that is distinct for the
extreme cases of a pure 2N or pure 6q  cluster.  We have presented
simple model estimates of the size of effects that may be seen,
showing that factors of two differences from maximum to minimum may
be expected in the 6q  case,  whereas no maximum to minimum
difference is expected in the 2N case.

{\underline{Acknowledgments}}. CEC thanks the National Science
Foundation for support under grant PHY-9112173  and KEL thanks the
Department of Energy for support under grant DOE No W-7405-ENG- 82
Office of Energy Research (KA-01-01).  We also thank Keith
Griffioen,  Jorge Morfin,  Brian Quinn,   and Mark Strikman for
useful remarks.

\newpage
\appendix\section{ Appendix: Distributions for six-quark clusters}

We apply the notation $|6q\rangle$ or 6q to label the situation when
the neutron and proton are melded  and lose their individual
identity.  This notation emphasizes the fact that standard QCD
quark  parton model considerations should be applied to this
interesting multiquark object.  Therefore,   for a generic
$|nq\rangle$ state  the sea,  valence,  and gluon distribution
functions (times $z$) are  written

\begin{eqnarray}
\overline {U}_n(z) &=& A_n (1-z)^{a_n} \nonumber \\
  V_n(z)    &=& B_n z^{1/2} (1-z)^{b_n} \nonumber \\
  G_n(z)    &=& C_n  (1-z)^{c_n},
\end{eqnarray} where $z$ is the fraction of the total cluster
momentum.  The coefficients and powers are  determined in
\cite{ls88} by appealing to standard normalization and momentum
conservation  considerations along with input information from
experimental study of the $n=2$ (pion) and  $n=3$ (nucleon)
situations.  As a result,  3 cases were developed to illustrate the
sensitivity to  small changes in the power of $(1-z)$,  i.e.,  $(a_6
, b_6) = (11,9)$ for case A;
$(11,10)$ for case B;   and $(13,10)$ for case C.

\begin{figure}

\vglue 4in  
\hskip 0.5in {\special{picture one scaled 1000}} \hfil

\caption{
  A putative ratio $R_1$ assuming the measured cross section
  for backward protons in deep inelastic scattering is dominated by
  6-quark configurations and using the LS model [9] for the
  distribution functions of the 6-quark cluster.  In general,  $R_1$
  is the measured cross section (sans some  kinematic factors)
  divided by the neutron structure
  function $F_{2n}$.    If backward  proton
  production were dominated by 2-nucleon correlations,  the plot
  would  be flat.  The plot is for fixed $\alpha = 1.4$
  (momentum 322 MeV for directly backward protons) and uses CH [10]
  distribution functions to
  obtain $F_{2n}$.  The heavy curve is from LS parametrization A,
  the normal
  curve is  from parametrization B,  and the dotted curve is from
  parametrization C.  The vertical units are arbitrary as the
  fragmentation function $D_{p/5q}$ is not normalized.}

   \label{fig1}
\end{figure}

\begin{figure}

\vglue 4in  
\hskip 0.5in {\special{picture two scaled 1000}} \hfil

\caption{Like Fig.\ 1 except that $F_{2n}$ is gotten from the CTEQ
  [11] distribution  functions,  specifically from the set CTEQ1L for
  $Q^2 = 4$ GeV$^2$.  The  heavy curve is from LS
  parametrization A, the normal curve is from  parametrization B,
  and the dotted curve is from parametrization C.   The curves turn
  up as $x \rightarrow  0.6 \ (\xi \rightarrow  1)$ because the CTEQ
  distribution  functions all approach zero at the upper limit faster
  than $(1-\xi)^3$,  and the
  fragmentation function we use goes to zero as $(1-z)^3$.}

   \label{fig2}
\end{figure}

\begin{figure}

\vglue 3.5in  
\hskip 0.5in {\special{picture three scaled 1000}} \hfil

\caption{The scaling window for $\alpha = 1.4$.  Values of $x$
  between the two  curves can be reached in the scaling region for a
  given incoming electron  energy $E$,  given in GeV above.  The
  lower curve is set by the requirement  that $Q^2 \geq 1$ GeV$^2$
  and the upper curve by the
  requirement that $W \geq 2$ GeV. The curves begin at $E = 3.72$ GeV
  for this $\alpha$.}

   \label{fig3}
\end{figure}


\newpage

\begin{thebibliography}{99}

\bibitem{fs79}  L.L. Frankfurt and M.I. Strikman,  Phys. Lett. B {\bf
69}, 93 (1977).

\bibitem{cls91}  C.E. Carlson,  K.E. Lassila,  and U.P. Sukhatme,
Phys. Lett. B {\bf 263},  277 (1991).

\bibitem{cl93}  C.E. Carlson and K.E. Lassila,  Phys. Lett. B {\bf
317}, 205 (1993).

\bibitem{mats89}  E. Matsinos et al.,  Z. Phys. C {\bf 44},  79
(1989).

\bibitem{ammo86}  V.V. Ammosov et al.,  Sov. J. Nucl. Phys. {\bf 43},
759 (1986).

\bibitem{efre80}  V.I. Efremenko et al.,  Phys. Rev. D {\bf 22},
2581 (1980).

\bibitem{berg78}  J.P. Berge et al.,  Phys. Rev. D {\bf 18},  1367
(1978).

\bibitem{atti91} C. Ciofi degli Atti,  S. Simula, L. L. Frankfurt,
and M. I. Strikman,  Phys. Rev. C {\bf 44}, R7 (1991).

\bibitem{ls88}  K.E. Lassila and U.P. Sukhatme,  Phys. Lett. B {\bf
209},  343 (1988).

\bibitem{ch83}  C.E. Carlson and T.J. Havens,  Phys. Rev. Lett. {\bf
51},  261 (1983).

\bibitem{cteq93}  J. Botts et al.,  Phys. Lett. B {\bf 304},  159
(1993).

\bibitem{mq86} A. Mueller and J.W. Qiu,  Nucl. Phys. B {\bf 268},
427 (1986); F. Close,  R. Roberts,  and J.W. Qiu,  Phys. Rev. D
{\bf 40},  2820 (1989).

\bibitem{scpv} M. Sato, S. Coon, H. Pirner and J. Vary, Phys. Rev. C
{\bf 33}, 1062 (1986).

\end{thebibliography}
\end{document}